\documentclass[11pt]{aastex}
\usepackage{emulateapj5}
\usepackage{psfig}
\usepackage{pslatex}
\begin{document}

\bibliographystyle{klunamed}

\submitted{Intl. Sp. Sci. Inst. Cosmic-Ray Conference Proceedings, 2000}

\title{The importance of anisotropic interstellar
turbulence and molecular-cloud magnetic mirrors 
for galactic cosmic-ray propagation}

\author{ B. D. G. Chandran\footnote{e-mail: benjamin-chandran@uiowa.edu} \\
\affil{Dept. Physics \& Astronomy, University of Iowa, Iowa City, IA}
}

\begin{abstract}

Recent studies suggest that when magnetohydrodynamic (MHD) turbulence
is excited by stirring a plasma at large scales, the cascade of energy
from large to small scales is anisotropic, in the sense that small-scale fluctuations
satisfy the inequality $k_\parallel \ll k_\perp$, where $k_\parallel$
and $k_\perp$ are, respectively, the components of a fluctuation's
wave vector $\parallel$ and $\perp$ to the background magnetic field.
Such anisotropic fluctuations are very inefficient at scattering
cosmic rays.  Results based on the quasilinear approximation for
scattering of cosmic rays by anisotropic MHD turbulence are presented
and explained. The important role played by molecular-cloud magnetic
mirrors in confining and isotropizing cosmic rays when scattering is
weak is also discussed.
\end{abstract}

\section{Introduction}

In diffusion models of Galactic cosmic-ray propagation,
cosmic rays are scattered by small-scale fluctuations in
the interstellar magnetic field. For cosmic-ray energies
below $\sim 10^2 -10^3$ GeV, these small-scale
fluctuations can arise from resonant waves that the 
cosmic rays generate themselves.
At higher energies, it is believed that self-confinement
is not possible, because the growth rates of the resonant
modes become too small in comparison to the rates at which
the modes are damped (\citeauthor{ces80} \citeyear{ces80},
\citeauthor{ber90} \citeyear{ber90}). For cosmic rays with energies
above $10^2-10^3$ GeV, scattering
can result from turbulence that is generated by large-scale
stirring of the interstellar medium (ISM), which results in a
cascade of magnetic energy from large to small scales.
Recent studies of magnetohydrodynamic (MHD) turbulence,
however, find that the small-scale fluctuations resulting
from a turbulent cascade satisfy the inequality $k_\parallel \ll k_\perp$,
where $k_\parallel$ and $k_\perp$ are, respectively, the components of a 
fluctuation's wave vector $\parallel$ and $\perp$ to the background
magnetic field ${\bf B}_0$ (e.g., \citeauthor{gol95} \citeyear{gol95}). 
This paper explains why fluctuations with this
anisotropy are very inefficient at scattering cosmic rays,
and presents results from quasilinear theory for the scattering
mean free paths resulting from anisotropic MHD turbulence.
The role of molecular clouds in confining and isotropizing
cosmic rays when
scattering is weak is also discussed.

\section{Anisotropic MHD turbulence and the Goldreich-Sridhar spectrum}

Early studies of MHD turbulence assumed that when
a plasma is stirred on some large scale $l$,
the cascade of energy from large scales to small scales
proceeds isotropically in ${\bf k}$-space (\citeauthor{kra65}
\citeyear{kra65}). More recent studies, however, find that energy cascades
efficiently to large values of $k_\perp$, but not
very efficiently to large values of $k_\parallel$
(e.g., \citeauthor{gol95} \citeyear{gol95}, \citeauthor{she83}
\citeyear{she83}).
(Note: if the mean-magnetic field is weaker
than the fluctuating magnetic field, then within
any stirring-scale cell of volume $l^3$ there is a preferential
field direction which can be thought of as the background
field ${\bf B}_0$ for all of the small-scale fluctuations within that
cell. Local anisotropy is then determined relative to the direction
of ${\bf B}_0$ within each stirring-scale cell.)

\citeauthor{gol95} (\citeyear{gol95}) have proposed an inertial-range
power spectrum for strong anisotropic MHD turbulence
in which the magnetic field fluctuations are comparable
to ${\bf B}_0$:
\begin{equation}
E_B(k_\perp, k_\parallel) \propto
k_\perp^{-10/3} l^{-1/3} g\left(\frac{k_\parallel}{
k_\perp^{2/3}l^{-1/3}}\right),
\label{eq:Ms} 
\end{equation} 
where the dimensionless function $g(x)$ is $\sim 1$ for $|x| \lesssim
1$ and rapidly approaches 0 for $|x| \gg 1$.
Evidence in support of equation~(\ref{eq:Ms})  has been
found in direct numerical simulations of MHD turbulence
(\citeauthor{mar00} \citeyear{mar00}, \citeauthor{cho00} \citeyear{cho00}). 
In this spectrum, there is only power at small scales when
$k_\parallel \lesssim k_\perp^{2/3} l^{-1/3}$.
In this region of ${\bf k}$-space, in which turbulence
is excited, the linear incompressible Alfv\'en-wave
period $(k_\parallel v_A)^{-1}$ is greater than the
nonlinear energy-transfer time $(k_\perp v_k)^{-1}$,
where $v_A = B_0/\sqrt{4\pi\rho}$ is the Alfv\'en speed,
$\rho$ is the mass density of the medium, and $v_k 
\sim v_A (k_\perp l)^{-1/3}$ is the rms velocity fluctuation
on a perpendicular scale of $k_\perp^{-1}$.
Because of this inequality, the Iroshnikov-Kraichnan
mechanism for slowing energy-transfer does not apply
(\citeauthor{kra65} \citeyear{kra65}). 

\section{Why scattering is weak in anisotropic turbulence}

If the magnetic power spectrum is isotropic in ${\bf k}$-space
or possesses slab symmetry (in which the wave vectors of fluctuations
are $\parallel$ to ${\bf B}_0$), then cosmic-ray scattering
is dominated by magnetostatic gyroresonant interactions, in which
the cosmic ray and fluctuation satisfy the resonance relation
\begin{equation}
k_\parallel v_\parallel = n \Omega,
\label{eq:msgr} 
\end{equation} 
where $v_\parallel$ is the component of a cosmic ray's velocity along
${\bf B}_0$, $n$ is a non-zero integer ($n= \pm 1$ for
slab symmetry), and $\Omega$ is the cosmic
ray's gyrofrequency. In the general resonance relation, the linear
wave frequency $\omega$
appears on the left-hand side of equation~(\ref{eq:msgr}).
It is neglected here since for incompressible Alfv\'en waves 
$\omega = k_\parallel v_A$, which is $\ll k_{\parallel} v_{\parallel}$
unless the angle between a cosmic-ray's
velocity vector ${\bf v} $ and ${\bf B}_0$,
the pitch angle, is very close to $90^\circ$.

If a cosmic ray's pitch angle isn't too close to $0^\circ$, 
$90^\circ$, or $ 180^\circ$, then
equation~(\ref{eq:msgr}) implies that the
fluctuations that dominate scattering satisfy $k_\parallel \sim \rho^{-1}$,
where $\rho = v_\perp / \Omega$ is a cosmic ray's gyroradius and $v_\perp$
is the component of ${\bf v}$  $\perp$ to ${\bf B}_0$. 
However, if $\rho^{-1} \gg l^{-1}$, where $l$ is the scale at which
the turbulence is stirred, and if the turbulence has the type of anisotropy
described by equation~(\ref{eq:Ms}), then the only fluctuations with
$k_\parallel \sim \rho^{-1}$ have $k_\perp \gg \rho^{-1}$, as depicted
in figure 1. But if $k_\perp \rho \gg 1$, then during
a single gyro orbit a cosmic ray traverses many uncorrelated turbulent
fluctuations of the required $k_\parallel$. The contributions from
these different fluctuations tend to cancel, resulting in 
highly inefficient scattering.

\begin{figure*}
\centerline{\psfig{figure=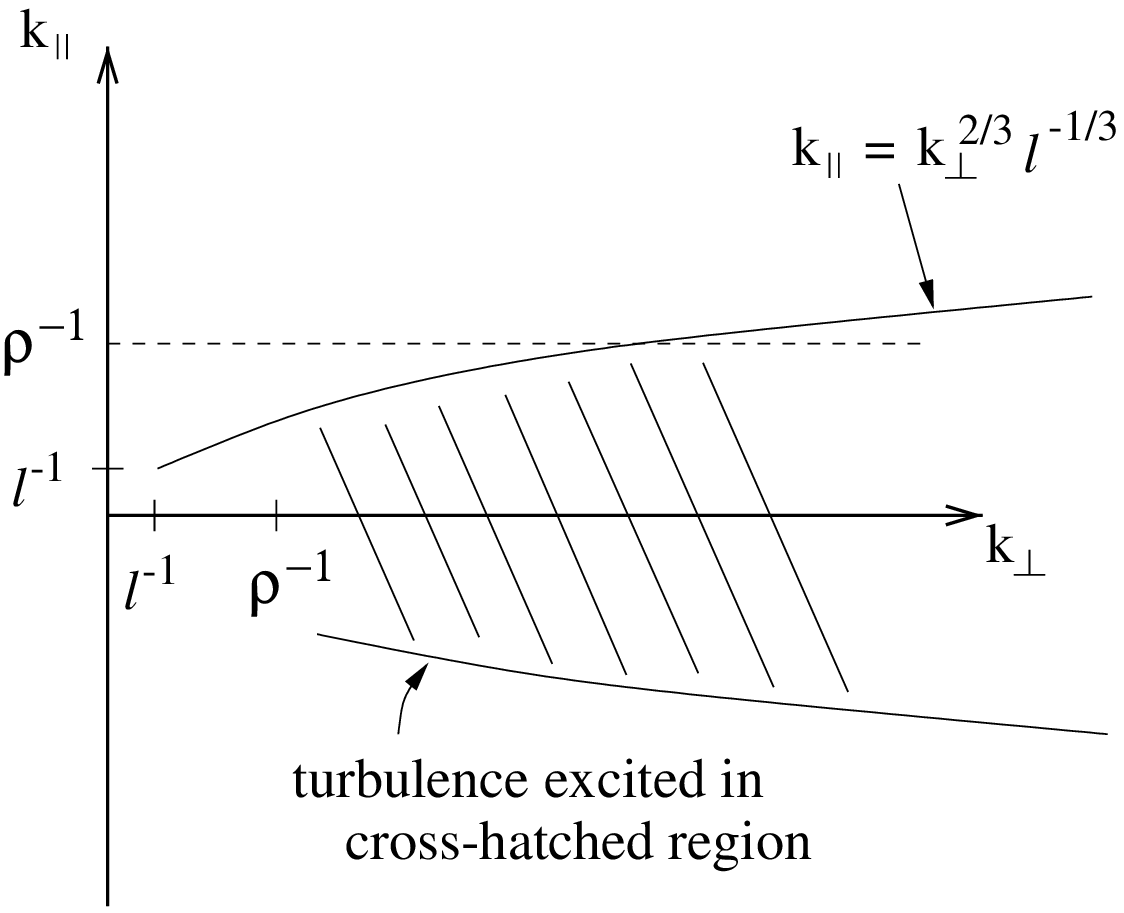,width=3in,clip=}}
\vspace{0.2cm}
\caption{}
\end{figure*}

Scattering rates have been calculated for turbulence with a power spectrum
described by equation~(\ref{eq:Ms}) in the quasilinear approximation
(\citeauthor{cha00a} \citeyear{cha00a}).
If
\begin{equation}
\delta \equiv \frac{v_A}{v} \ll 1,
\end{equation}
and 
\begin{equation}
\epsilon \equiv \frac{v}{l\Omega} \ll 1,
\end{equation}
then when  $\epsilon^{3/2} \ll (-\ln \epsilon)\delta $
the coefficient of spatial diffusion along the magnetic field resulting
from the quasilinear scattering rates is given by  (\citeauthor{cha00a}
\citeyear{cha00a})
\begin{equation}
\kappa_\parallel
= v l (-\delta \ln \epsilon)^{-1}\left(\frac{5}{2} - \frac{3\pi}{4}\right).
\label{eq:kappap} 
\end{equation} 
This value is far too large to explain confinement of cosmic
rays to the Galaxy. Thus, some mechanism besides turbulence
described by equation~(\ref{eq:Ms})  must be invoked to
explain cosmic-ray confinement. For cosmic rays with
energies less than $10^2 - 10^3$ GeV, waves excited by 
the cosmic rays may provide the confinement. For higher
energy cosmic rays for which self confinement does not appear
possible (\citeauthor{ces80} \citeyear{ces80}, \citeauthor{ber90}
\citeyear{ber90}), molecular-cloud magnetic
mirrors may play an important role. It should be noted that
equation~(\ref{eq:kappap}) applies to both cosmic-ray 
nuclei and electrons.

\section{Why molecular clouds can help confine cosmic rays
when scattering is weak}
\label{sec:clouds} 

Molecular-clouds are characterized by a range of sizes, masses, and
densities.  Cloud mass spectra obey power-law
scalings over several decades of cloud masses, and a power-law
mass-size relation also holds over a range of scales (\citeauthor{elm96}
\citeyear{elm96}, \citeauthor{hei98} \citeyear{hei98}, \citeauthor{bli97}
\citeyear{bli97}).
\citeauthor{elm97} (\citeyear{elm97})
has proposed a useful quasi-fractal model for this
hierarchy of structure, with smaller, denser objects nested within
structures that are larger and more diffuse. Using straightforward
rules to generate fractal structures, Elmegreen estimates through
numerical modeling that a line-of-sight through a fractal molecular
cloud complex has a $ 50\pm 10\%$ chance of entering dense molecular
material, and a $ 50\pm 10\%$ chance of passing through a hole in the
fractal complex filled with diffuse matter.  Elmegreen interprets the
standard 8 large absorption lines per kpc (\citeauthor{bla52}
\citeyear{bla52}) as evidence for
an average of 3 fractal cloud complexes per kpc along a typical line
of sight. Because of the 50\% see-through probability, however,
photons can travel $\sim 600$ pc without entering molecular material.
The 8 absorption lines per kpc are then not spaced at even distances
along a line of sight, but rather are clustered in groups of $\sim 5$
per cloud complex.

\begin{figure*}
\centerline{\psfig{figure=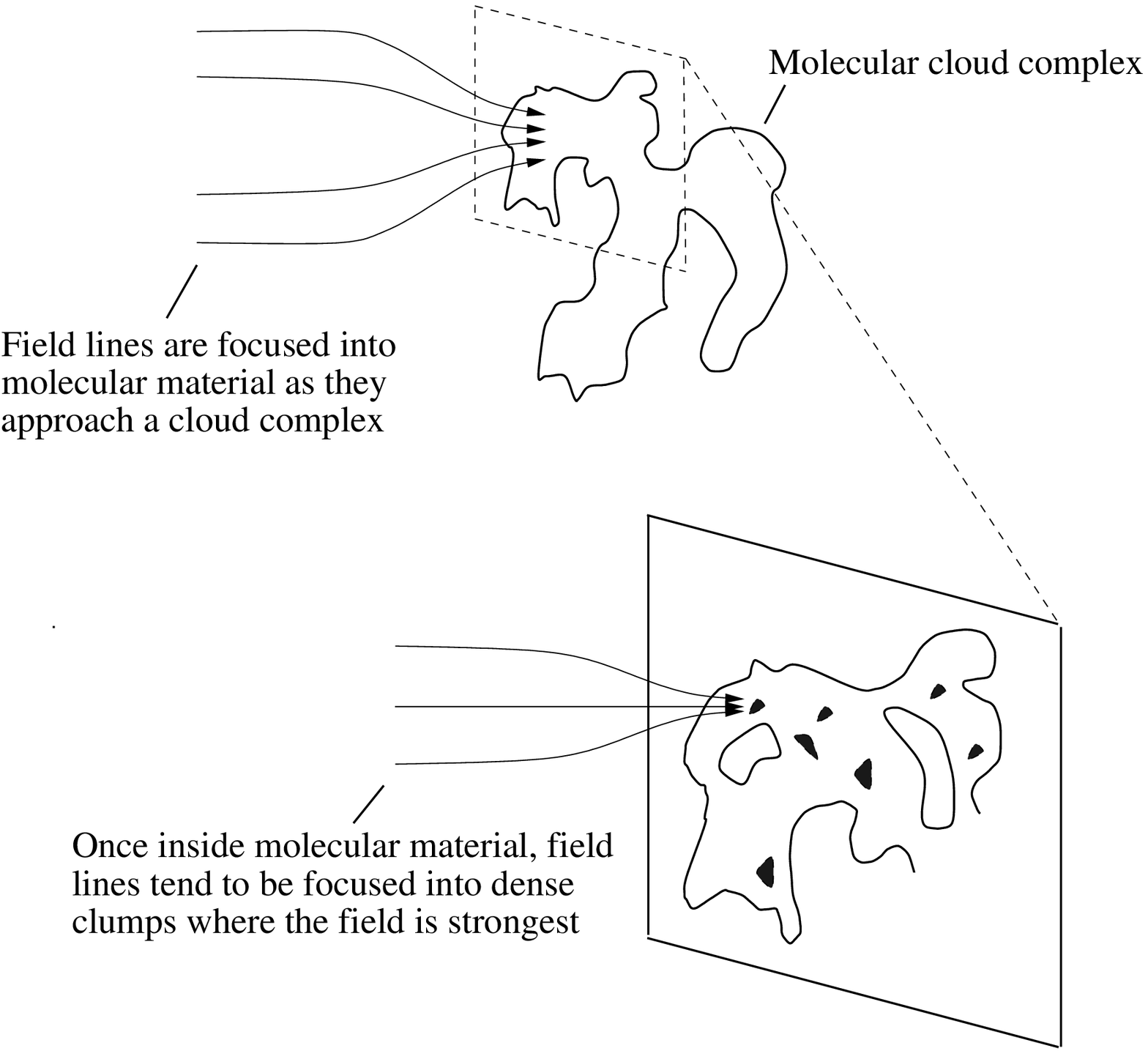,width=4in,clip=}}
\vspace{0.2cm}
\caption{\label{fig:combined_focus} }
\end{figure*}

Because magnetic field lines are focused into strong-field regions,
and because the magnetic field is stronger within molecular clouds
than in the ICM, a magnetic field line passing
through a cloud complex has a higher probability of entering molecular
material than a straight line of sight (figure~2). If
the mean field strength within the molecular material is $m$ times the
typical field strength in the ICM, and if 50\% of the
lines-of-sight through the complex intersect molecular material, then
one would expect a fraction $P$ on the order of $m/(m+1)$ of the magnetic flux (and,
therefore, field lines) through the complex to pass through molecular
material. The line-of-sight average of a cloud's magnetic field can be
obtained through Zeeman splitting.  \citeauthor{tro86} (\citeyear{tro86}) report
Zeeman measurements of molecular-cloud field strengths ranging from 9
$\mu$G to 120 $\mu$G. The typical field strength in the
ICM is $\sim 4- 5$ $\mu$G (\citeauthor{zwe97} \citeyear{zwe97}).
The ratio $m$ is thus fairly large, and $P$ is close to 1.  Moreover, due
to magnetic focusing, the spacing of cloud complexes along field lines
may be smaller than the $\sim 300$-pc spacing of complexes along
straight lines. On the other hand, tangling of field lines in the
ICM may tend to increase the spacing of successive cloud 
complexes along a given field line as measured along that field line.
It seems reasonable, however, to take the typical distance between
cloud complexes as measured along a field line to be
\begin{equation}
l_{\rm intercloud}  \simeq 300 \mbox{ pc}.
\end{equation} 

Once inside a molecular cloud, field lines tend to be focused into dense
clumps where the field strength is larger than the average field strength
in the cloud. A range of observations suggests that at particle densities $n$
above $10^2 \mbox{ cm}^{-3}$, the magnetic field strength $B$ scales
as (\citeauthor{val97} \citeyear{val97})
\begin{equation}
B\propto n^{0.5}.
\label{eq:Bn}
\end{equation}   
According to \citeauthor{hei98} (\citeyear{hei98}),
\begin{eqnarray} 
M & \propto & r^{2.3}\label{eq:Mr},  \\
n & \propto & r^{-0.7} \label{eq:nr}, \mbox{ and} \\
dN/dr & \propto & r^{-3.0} \label{eq:dNdr}, 
\end{eqnarray} 
where $M$ is the mass of a clump of linear dimension $r$, and $(dN/dr) \triangle r$ is the
number of clumps with linear dimension $r$ in the interval $(r, r+\triangle r)$.
Since the flux through a clump is $\sim Br^2$, 
equations~(\ref{eq:Bn}) through (\ref{eq:dNdr}) imply that
the total flux $\Phi$ through clumps with linear dimension between $r$ and $2r$ 
scales as 
\begin{equation}
\Phi \propto r^{-0.35}.
\end{equation} 
That is, there is more flux through the smaller, denser clumps than
through the cloud complex as a whole. This means that as a single
field line passes through a complex, it must on average pass through
several of the densest clumps described by the power-law scalings,
clumps in which the magnetic field strength is large.

If $B_{\rm ICM}$ is the field strength in the ICM and $B_{\rm max}$
is the maximum field strength encountered by a cosmic ray in
a molecular-cloud complex, 
then the cosmic ray will be magnetically
reflected by the cloud complex provided that its
pitch-angle cosine $\xi_{\rm ICM} = v_\parallel/v$ in the ICM as it
approaches the complex satisfies the inequality
\begin{equation}
|\xi_{\rm ICM}| < \sqrt{1 - \chi^{-1}},
\label{eq:trapcond}
\end{equation}  
where
\begin{equation}
\chi \equiv \frac{B_{\rm max}}{B_{\rm ICM}}.
\end{equation} 
The above discussion suggests that $\chi \gg 1$,
and thus molecular clouds can magnetically reflect a large fraction
of cosmic rays.

\begin{figure*}
\centerline{\psfig{figure=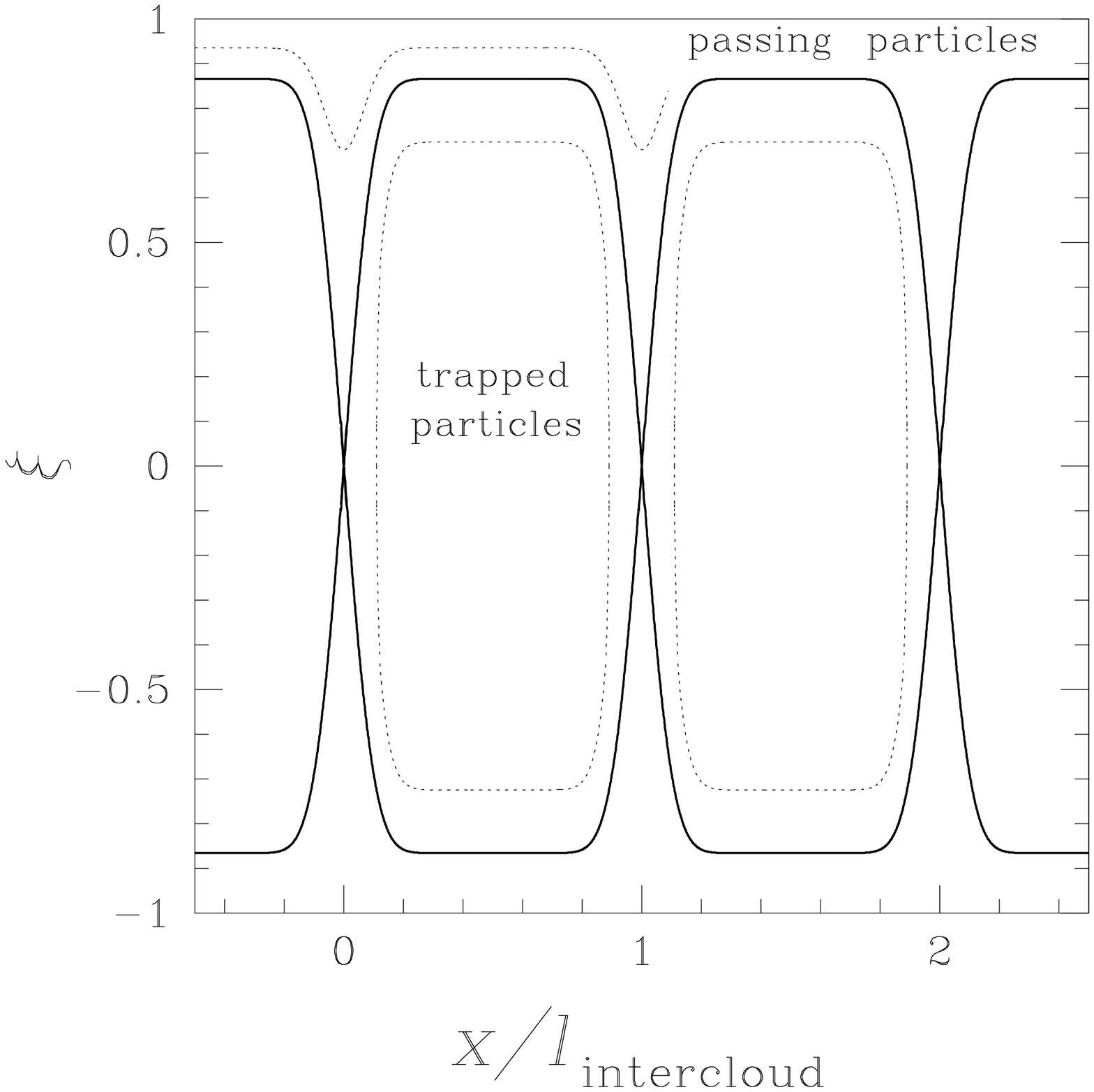,width=3.5in,clip=}}
\vspace{0.2cm}
\caption{\label{fig:phase} }
\end{figure*}

When the scattering mean-free path $ \kappa_\parallel /v$ is
much greater than $l_{\rm intercloud}$, cosmic rays travel between
molecular clouds without significant scattering. Phase-space
orbits of cosmic rays in this weak-scattering limit are depicted
in figure 3, where $\xi = v_\parallel /v$.
Particles with $|\xi|$ close to 1 are not magnetically reflected
and can pass through molecular clouds. Trapped cosmic rays move
on closed orbits in the $x$-$\xi$ plane, where $x$ is distance along
a field line. 

The way in which molecular-cloud magnetic mirrors affect cosmic-ray
transport depends upon cosmic-ray energy and also the efficiency
of cosmic-ray scattering. Approximate coefficients of diffusion
perpendicular to the Galactic plane have been calculated for several
different propagation
regimes (\citeauthor{cha00b} \citeyear{cha00b}).
Broadly speaking, cosmic rays can escape from
magnetic traps in one of two ways, by scattering into the
passing region of phase space and then traveling
along the magnetic field through a molecular cloud, or by drifting perpendicular
to the magnetic field.  The second case is depicted in
figure 4: a cosmic ray initially trapped between clouds A
and B can drift perpendicular to the magnetic field and
end up trapped between clouds A and D.

Because molecular clouds are confined to the Galactic disk,
it is possible that they have no affect on cosmic-ray propagation
in the halo. On the other hand, if the magnetic field lines in the
halo possess numerous arcs that are anchored down to the
disk on either end (analogous to closed field lines in the solar
corona), then molecular cloud magnetic mirrors may
affect propagation in the halo to some extent.

\begin{figure*}
\centerline{\psfig{figure=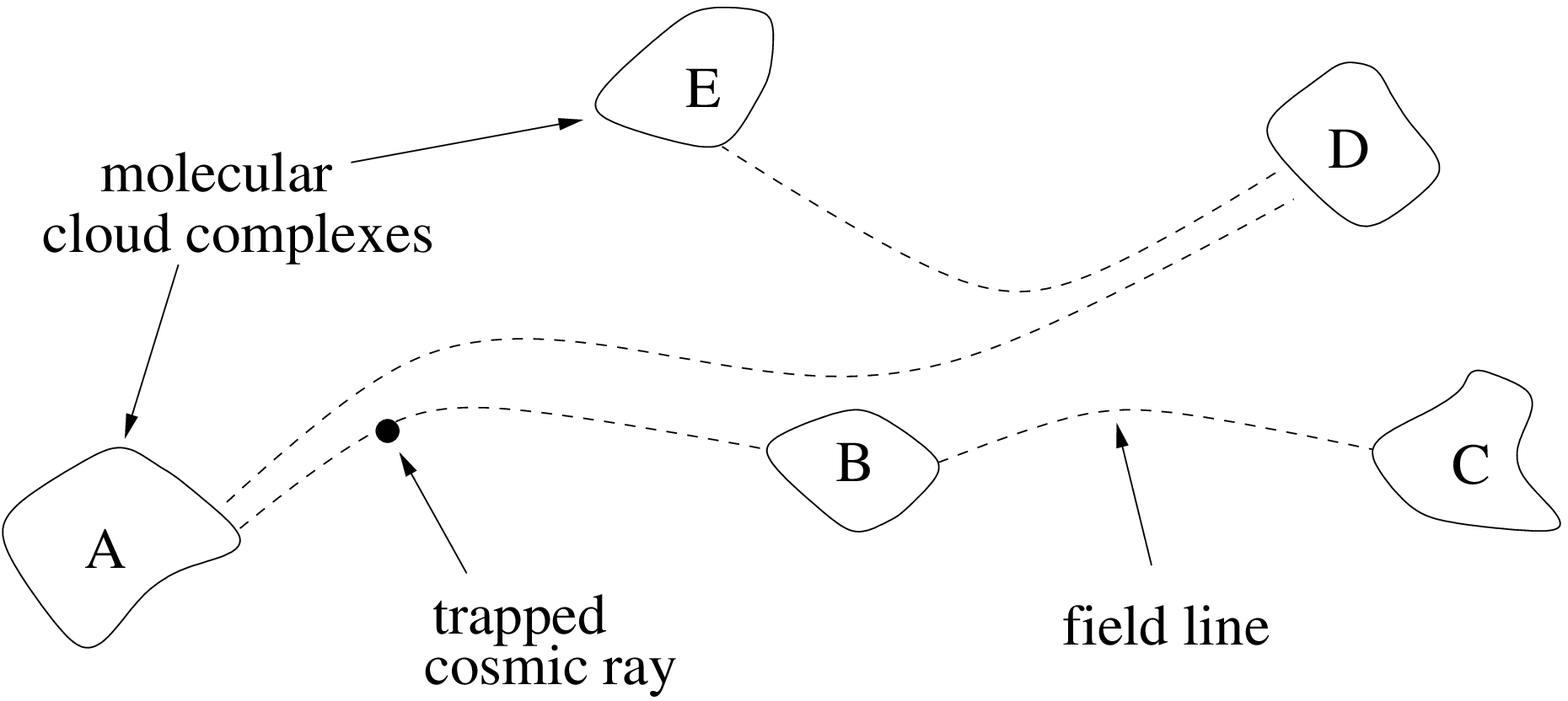,width=4in,clip=}}
\vspace{0.2cm}
\caption{\label{fig:trapped_diffusion} }
\end{figure*}

\section{Isotropization of cosmic rays by molecular-cloud magnetic mirrors}

One of the sources of cosmic-ray anisotropy is the flow of cosmic rays
along the magnetic field. Because molecular-cloud magnetic mirrors
impede this flow, they reduce the level of anisotropy for any given level
of weak scattering. Values of the harmonics of the cosmic-ray distribution
function (as functions of the scattering rate, $\chi$, 
and $l_{\rm intercloud}$)
are given by \citeauthor{cha00b} (\citeyear{cha00b})
under the assumption that the pitch-angle
scattering frequency is weak and independent of pitch angle.

\section{Implications of molecular-cloud magnetic mirrors for diffuse
gamma radiation and secondary products}

When scattering is weak, the density of cosmic rays within molecular
clouds $n_{\rm cloud}$ is determined by two competing effects. On the
one hand, cosmic rays are reflected as they approach cloud complexes,
which tends to reduce $n_{\rm cloud}$.  On the other hand, magnetic
field lines are brought closer together in high-field regions, which
acts to increase $n_{\rm cloud}$ since cosmic rays travel primarily
along the magnetic field.  It can be shown that when energy losses are
neglected, these two effects cancel (\citeauthor{cha00b} \citeyear{cha00b}).
This point is important since 
if $n_{\rm cloud}$ were in fact less than $n_{\rm
ICM}$, there would be a corresponding reduction in spallation and
diffuse gamma radiation for a fixed average energy density of cosmic
rays throughout the Galaxy. (For sufficient
ionization losses at low cosmic-ray energies, it should be noted
that the value of
$n_{\rm cloud}$ can be reduced below the cosmic-ray density in the
intercloud medium $n_{\rm ICM}$.)

The two competing effects described above are illustrated graphically
in figure~5. Each of the two narrow flux tubes in
figure~5 has a bounding surface that is everywhere
parallel to the magnetic field. The cross-sectional area of each tube
is proportional to $(1/B)$, where $B$ is the
field strength. Since motion perpendicular to the magnetic
field is suppressed, the cosmic rays within each flux tube to a good
approximation remain within their respective flux tubes as they move
along the field. Because of magnetic mirroring, the number of cosmic
rays per unit length within a flux tube decreases in high-field
regions.  However, because the cross-sectional area of the flux tube also decreases,
the number of cosmic rays per unit volume stays the same.

\begin{figure*}
\centerline{\psfig{figure=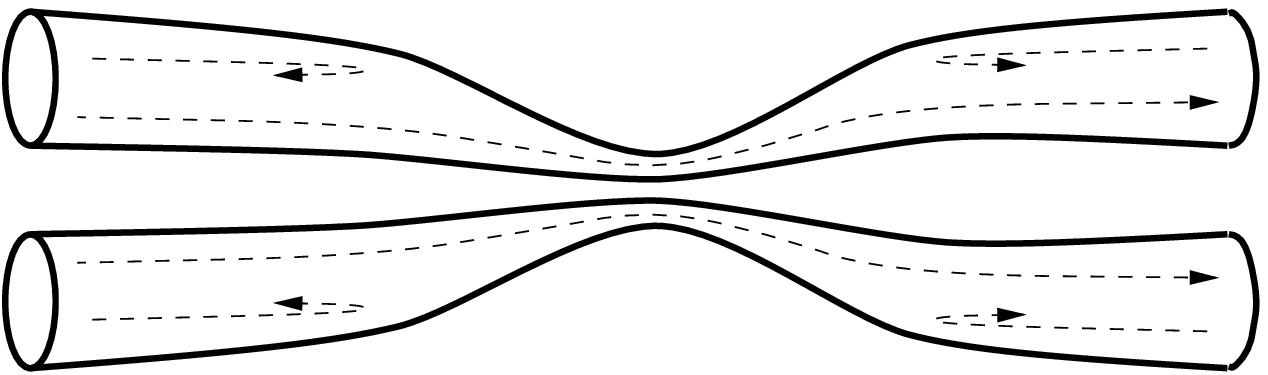,width=4in,clip=}}
\vspace{0.2cm}
\caption{\label{fig:tubes} }
\end{figure*}

\section{Does the model fit the data?}

At this stage, it is difficult to determine from observations whether molecular
clouds play a role in cosmic ray confinement. Although
the observed energy dependence of the cosmic-ray path length at
cosmic ray energies $< 10^2$ GeV provides important information
on propagation at energies $< 10^2$ GeV, almost nothing is known
about the path length at the energies above $10^2-10^3$ GeV
at which self-confinement appears to break down and at which confinement
may depend upon molecular clouds. There appear to be three main 
possibilities. First, molecular clouds may 
help confine cosmic rays at energies above $10^2-10^3$ GeV
as described in this paper. Second, the arguments that self-confinement
breaks down at energies above $10^2-10^3$ GeV may be incorrect. Third, 
interstellar turbulence generated
by large-scale stirring may possess features not described by the
Goldreich-Sridhar theory that allow for stronger scattering.

\section{Conclusion}

Recent investigations into MHD turbulence are providing
new and important results on the anisotropy of the small-scale
fluctuations that result from a cascade of magnetic energy
from large to small scales. As discussed in this paper,
anisotropic small-scale fluctuations
are inefficient at scattering cosmic rays. For cosmic
rays with energies less than $10^2- 10^3$ GeV,
resonant waves excited by streaming cosmic rays are believed
to be sufficient to confine cosmic rays to the Galaxy regardless of the
nature of the cascade in MHD turbulence. At higher energies,
however, it is believed that such self-generated waves
are insufficient. Thus, if scattering by the turbulence that
is generated by large-scale stirring of the ISM is inefficient,
then some additional mechanism is needed to confine 
and isotropize
cosmic rays at energies above $\sim 10^2 - 10^3$
GeV. Such a mechanism may be provided by molecular-cloud
magnetic mirrors.

\end{document}